\documentclass[aps,amsmath,amssymb,reprint,twocolumn]{revtex4-2}
\usepackage[dvipsnames]{xcolor}
\usepackage{graphicx}
\usepackage{physics}
\usepackage{enumitem}
\usepackage{todonotes}
\usepackage{hyperref}
\usepackage[capitalize]{cleveref}

\newcommand{\bs}[1]{\boldsymbol{#1}}

\begin{document}

\title{Geometric control of motility-induced phase separation}

\author{Toler H. Webb}
\thanks{These authors contributed equally to this work}
\affiliation{Department of Physics, Emory University, Atlanta, GA 30322, USA}

\author{Helen S. Ansell}
\thanks{These authors contributed equally to this work}
\affiliation{Department of Physics, Emory University, Atlanta, GA 30322, USA}

\author{Daniel M. Sussman}
\email{daniel.m.sussman@emory.edu}
\affiliation{Department of Physics, Emory University, Atlanta, GA 30322, USA}

\begin{abstract} 
Curvature fundamentally alters the collective properties of soft, active, and biological materials.
Here we study motility-induced phase separation (MIPS), a canonical non-equilibrium transition, and demonstrate that even weak and slowly varying curvature provides robust geometric control over the dense MIPS phase.
This includes dictating both the location and morphology of the MIPS cluster, even in regimes where the effect on the overall phase boundaries is minimal.
Focusing on active Brownian particles confined to the surface of a torus, we show that varying the aspect ratio drives a structural transition of the dense cluster from a disk localized at the outer equator to a band wrapping the minor circumference.
We then discuss how the curved geometry provides a platform for comparing different theoretical frameworks for the MIPS phase: by analyzing the geometries of the cluster boundaries, we compare the structures predicted by thermodynamic and kinetic pictures.
Our results establish curved space not only as a tool to shape and guide non-equilibrium dynamics, but as a uniquely sensitive arena for probing the fundamental mechanisms of active matter.
\end{abstract}

\maketitle

\section{Introduction}

Curvature fundamentally alters the properties of materials and is an inherent feature of many biological environments. 
In equilibrium soft matter, curvature dictates the geometry and dynamics of phase separation~\cite{fonda2018interface,marenduzzo2013phase,busuioc2020axisymmetric,fonda2019thermodynamic,vandin2016curvature,horsley2018aspects,gomez2015phase,law2018nucleation,law2020phase}. 
In living systems, this geometric influence scales from the localized distribution of molecules in cell membranes~\cite{arnold2025bending} up to the curvotactic guidance of migrating cell collectives~\cite{yevick2015architecture,xi2017emergent,tang2022collective,pieuchot2018curvotaxis}. 
Motivated by the link between geometry and biological function, a growing body of work has sought to understand how synthetic active materials respond to curved environments. 
This has revealed that introducing spatial curvature to active systems yields novel phenomena, including modified flocking behavior~\cite{sknepnek2015active,shankar2017topological,hueschen2023wildebeest}, the trapping of semi-flexible filaments~\cite{janzen2026active}, coordinated continuous rotation~\cite{praetorius2018active,happel2022effects, happel2024coordinated,ai2020binary}, and curvature-dependent rigidity in epithelial models~\cite{sussman2020interplay,demarzio2025epithelial}.

Motility-induced phase separation (MIPS), in which self-propelled particles phase separate into coexisting dense and gaseous phases, is one of the most well-studied phenomena to emerge in scalar active matter systems~\cite{cates2015motility,obyrne2023introduction,cates2025active}.
This phenomenon is observed in computational studies of minimal active models \cite{fily2012athermal,redner2013structure,fily2014freezing,bialke2015negative}, in experimental studies of self-propelled colloids~\cite{buttinoni2013dynamical,vanderlinden2019interrupted}, and in collective motion of biological systems such as bacteria~\cite{liu2019self,ridgway2023motility} and fire ants~\cite{anderson2022social}.
The introduction of curvature to these models, whether through boundary confinement~\cite{knippenberg2024motility} or continuous surface geometry~\cite{iyer2023dynamics,schonhofer2022curvature}, fundamentally alters MIPS phase behavior.
Simulations of active Brownian particles (ABPs) have shown that surface curvature can alter the non-equilibrium phase diagram, with large curvature scales able to substantially alter the parameter regime in which phase separation occurs~\cite{iyer2023dynamics,schonhofer2022curvature}.
Sch{\"o}nh{\"o}fer and Glotzer, in particular, showed that positive curvature can promote clustering while negative curvature inhibits it, with the strength of this effect depending on the relative magnitude of the Gaussian curvature $K$ to the size scale of the individual particles $\sigma$ \cite{schonhofer2022curvature}. 
At lower curvature scales, for which the quantity $\abs{K}\sigma^2 \ll 1$, the phase boundaries approach the Euclidean space limit, as would be expected.

However, whether slowly varying curvature in this regime can exert control over the \emph{morphology} of the dense phase remains an open question. 
Here we show that it indeed has important consequences, even in parameter regimes where $\abs{K}\sigma^2$ is small and its effect on the overall phase boundary is minimal. 
By studying MIPS on the surface of a torus, which contains regions of positive and negative Gaussian curvature, we demonstrate that curvature provides robust geometric control over both the shape and location of the dense cluster.
Reminiscent of how equilibrium phases seek optimal perimeter-to-stress balances on curved surfaces~\cite{law2020phase}, at a low torus aspect ratio $\xi$, the dense active phase forms a disk that preferentially sits near the outer equator in the region of positive curvature.
At high $\xi$ the cluster instead forms a band that wraps the minor circumference of the torus.

Our curved space simulations also provide a platform for breaking the geometric degeneracy between different mechanisms that can drive the formation and shape of the dense cluster.
Many features of MIPS map directly onto expectations for equilibrium interfaces~\cite{solon2015pressure,cates2015motility,patch2018curvature}, and theoretical models suggest that cluster interfaces follow surface-tension-driven area-minimizing principles~\cite{langford2024theory,li2025surface}.
An alternative interpretation treats the cluster as a kinetic structure that is shaped by the balance of particle deposition and escape~\cite{tailleur2008statistical,redner2013structure, soto2014run, soto2024kinetic}.
In Euclidean space, both of these interpretations lead to the averaged dense phase cluster shape being a disk or a band, depending on the area fraction covered by the dense phase and the aspect ratio of the system~\cite{bialke2015negative}.
However, in regions of varying curvature, these interpretations predict distinguishable shapes for the cluster boundaries.
By leveraging the geometry of the torus, we directly compare the dense phase boundaries against the structures predicted by thermodynamic perimeter-minimization and kinetic frameworks.
Our results highlight how curvature can act as a powerful tool for revealing the physical principles that govern active processes.

\section{Methods}
We investigate active Brownian particles (ABPs) with soft repulsive interactions that move and interact entirely within a curved surface.
The motion of the particles is governed by the coupled overdamped Langevin equations
\begin{equation}
\begin{split}
    \bs{\dot r}_i &=  \bs{v}_i + \mu \bs{F}_i\\
    \dot\theta_i &= \eta^r_i, 
\end{split}
\end{equation}
where $\bs{r}_i$ is the position of particle $i$, $\theta_i$ is the orientation of its heading measured relative to a reference axis, and dots denote time derivatives.
The translational motion of the particles has contributions from the self-propulsion $\bs{v}_i$ and the net force $\bs{F}_i$ on a particle due to neighbor interactions.

The self-propulsion is given by $\bs{v}_i = v_0(\cos{\theta_i}\,\vu{e}_1 +  \sin{\theta_i}\,\vu{e}_2)$, where $v_0$ is the self-propulsion speed and $\vu{e}_1$ and $\vu{e}_2$ are orthogonal coordinates in the local tangent plane at a given point on the surface.
We set $v_0$ and the time-scale $\tau$ such that a free particle self-propels one particle diameter $\sigma$ in time $\tau$, giving $v_0=\sigma/\tau$, and we use $\dd t = 0.01\tau$ as our simulation timestep.
Particle orientations are subject to Gaussian white noise obeying $\expval{\eta_i^r}=0$ and $\expval{\eta_i(t)\eta_j(t')} = 2 D_r\delta_{ij}\delta(t-t')$. 
Here $D_r$ is the rotational diffusion constant, which has a corresponding timescale $\tau_r = 1/D_r$.

Particles interact with each other to give a net force $\bs{F}_i = \sum_j \bs{F}_{ij}$ on a particle due to its neighbors. 
Here we use a soft harmonic repulsion
\begin{equation}
    \bs{F}_{ij} = 
    \begin{cases}
    -k(\sigma-\abs{\bs{r}_{ij}})\bs{\hat r}_{ij}, & \text{if } \abs{\bs{r}_{ij}}<\sigma\\
    0, & \text{otherwise,}
    \end{cases}
\end{equation}
where $\bs{\hat r}_{ij}$ is a unit vector tangent to the geodesic connecting points $i$ and $j$ at point $i$, and $k$ is the stiffness.
We set $k=30$ and the mobility $\mu=1$ such that the elastic relaxation time scale $\tau_{e} = (\mu k)^{-1}$ is small.

Due to challenges posed by curvature, many studies of particulate systems on curved surfaces calculate interparticle forces and particle motion using a Euclidean metric with additional forces imposed to ensure the particles remain within the surface~\cite{law2018nucleation,law2020phase,schonhofer2022curvature,ai2020binary,janzen2026active,sknepnek2015active}.
Here, we instead use a fully curved-space approach where all particle motion and interactions occur within the surface using geodesic distances to calculate forces and in-surface vector transport for particle translation, allowing us to truly probe the effects of inherent curvature in this system.
We conduct all simulations using curvedSpaceSim~\cite{webb2025curvedspacesim}, which performs these calculations on arbitrary connected non-self-intersecting embedded surfaces represented by triangulated mesh objects.
The equations of motion are integrated using a geodesic-based Euler-Maruyama scheme, with rotational noise applied in the tangent plane of the particles.

In Euclidean space, the phase behavior of this system can be characterized in terms of the rotational P\'eclet number $\mathrm{Pe}_r = \frac{v_0}{\sigma D_r}$, which characterizes how far a particle moves before it changes orientation, and the total area fraction $\phi$ of particles on the surface~\cite{redner2013structure,fily2014freezing}.
On curved surfaces, the scale of the Gaussian curvature of the surface relative to the particle size then shifts the location of the phase boundaries~\cite{schonhofer2022curvature}.
Here we highlight the effect of curvature on properties of the dense phase in regimes where it has minimal effect on the overall phase separation. 
We therefore set $D_r=1/100$ to give $\mathrm{Pe}_r = 100$, and investigate $\phi = 0.4$, which sits well within the region of phase space where MIPS is expected to occur~\cite{redner2013structure,fily2014freezing,schonhofer2022curvature}.
We treat particles as circular disks with the diameter $\sigma$ determined such that a system of $N$ particles obeys $N\pi(\sigma/2)^2= \phi A $, where $A$ is the total surface area of the system.
The approximation that the area of a particle is its Euclidean area is reasonable for sufficiently large $N$, as used here. 
We verified that the phase behavior of our system is consistent with prior expectations by investigating interaction strengths between $k=10$ and $k=100$ and area fractions between $0.2$ and $0.8$ for a torus with aspect ratio $\xi=1.5$.

\begin{figure}
    \centering
    \includegraphics[width=\linewidth]{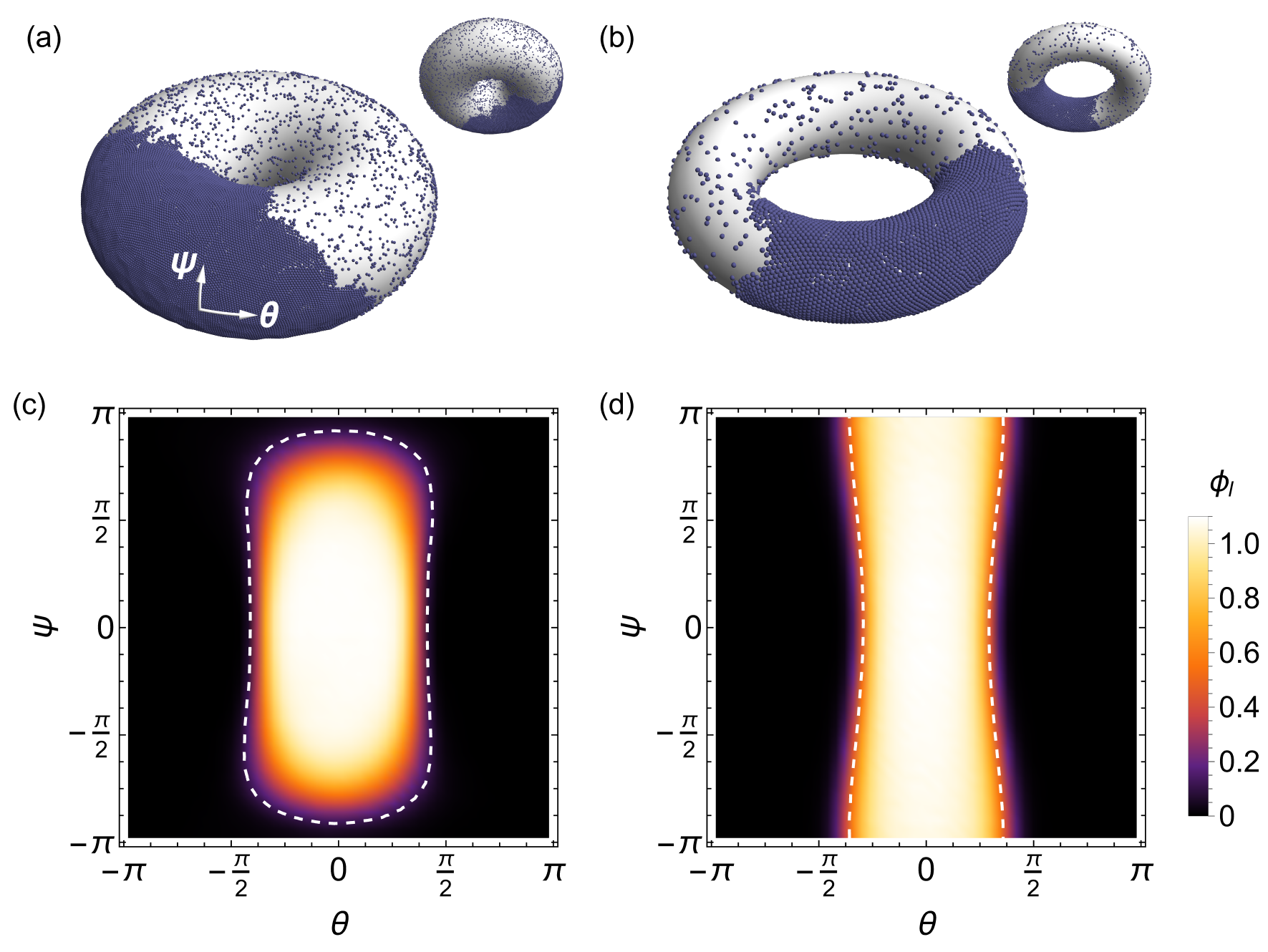}
    \caption{
    (a) Snapshot of a steady-state configuration for a torus with aspect ratio $\xi=1.5$ and $N=20,000$ particles. Here the dense phase forms a disk.
    (b) Steady-state snapshot for $\xi=3.0$ and $N=5000$. Here the dense phase forms a band that wraps around the minor circumference of the torus.
    (c-d) Corresponding time-averaged density profiles for particles in the dense phase. Cluster locations are centered in the $\theta$ direction before averaging. Dashed lines show the contour enclosing the average dense phase area fraction $\phi_d^*=0.314$. The outer equator has $\psi=0$.
    }
    \label{fig:schematics-density}
\end{figure}

Having established this baseline, we focus on how the specific geometric parameters of the torus influence MIPS.
Coordinates in the surface are parameterized by 
\begin{equation}
    \bs{r} = 
    \begin{pmatrix} x\\y\\z\end{pmatrix} = 
    \begin{pmatrix}
        (R+r\cos{\psi})\cos{\theta}\\
        (R+r\cos{\psi})\sin{\theta}\\
        r \sin{\psi}
    \end{pmatrix}
    \label{eq:r-torus}
\end{equation} 
where the major axis $R$ is the distance from the center of the torus to the center of the tube, and the minor axis $r$ is the radius of the tube. 
The toroidal angle $\theta$ is measured around the major circumference, while angles around the minor circumference are measured using the poloidal angle $\psi$ (see \cref{fig:schematics-density}a).
The torus aspect ratio is $\xi = R/r$, and we fix the overall system size by setting $r = 1/R$ so that the torus surface area $A = 4\pi^2 r R = 4\pi^2$ is constant. 
We vary the aspect ratio in the range $1.1\leq\xi\leq 3$ and study the resulting dense phase behavior.

The Gaussian curvature of the torus,
\begin{equation}
K = \frac{\cos{\psi}}{1 + \xi^{-1} \cos{\psi}},
\end{equation}
attains a maximum positive value at the outer equator ($\psi=0$) and a minimum negative value at the inner equator ($\psi=\pi$).
To ensure we are probing regimes where the curvature has minimal effect on the phase behavior, we set $\sigma$ to be small relative to the curvature scale.
Ref.~\cite{schonhofer2022curvature} identified that this low curvature regime is bounded by $\abs{K}\sigma^2\lesssim 0.003$, which sets a $\xi$-dependent lower bound on our choice of $N$.
At the outer equator, this varies from $N\approx3500$ for $\xi=1.1$ up to $N\approx5000$ for $\xi=3.0$.
At the inner equator, the magnitude of the curvature is larger, meaning we would require $N\approx 74000$ for $\xi=1.1$, which comes down to $N\approx 10000$ for $\xi=3$.
Given the computational intensity of our simulations, here we use $N=5000$ for the majority of our simulations exploring the behavior of the dense phase as a function of $\xi$. 
To verify that our results are robust at larger sizes, we have performed simulations with $N=20,000$ for select values of $\xi$ that span the different regimes we observe.

In all simulations, we initialize the system in a random configuration and allow the dynamics to reach steady state.
We then record snapshots spaced at intervals of at least $200\tau$ for a duration of at least $40,000\tau$. 
Full details of the data collected for each $\xi$ and $N$ are listed in \cref{tab:raw-data}.

\section{Results}

\subsection{Morphological transition and localization on the torus}

We begin by investigating general features of the dense phase as the aspect ratio $\xi$ of the torus is varied.
In the low $\xi$ regime, the dense phase is typically disk-shaped, while at high $\xi$ the dense cluster instead spans the torus minor circumference to form a band.
Typical snapshots in each of these regimes are shown in \cref{fig:schematics-density}(a--b) for $\xi=1.5$ and $3$ respectively.
To investigate the typical cluster shape, we plot the averaged density profiles for these $\xi$ values in \cref{fig:schematics-density}(c--d). 
These are measured after centering the clusters in each snapshot in the $\theta$ coordinate direction, and highlight the distinct disk and band shapes in the low- and high-$\xi$ regimes.

\begin{figure*}
    \centering
    \includegraphics[width=\textwidth]{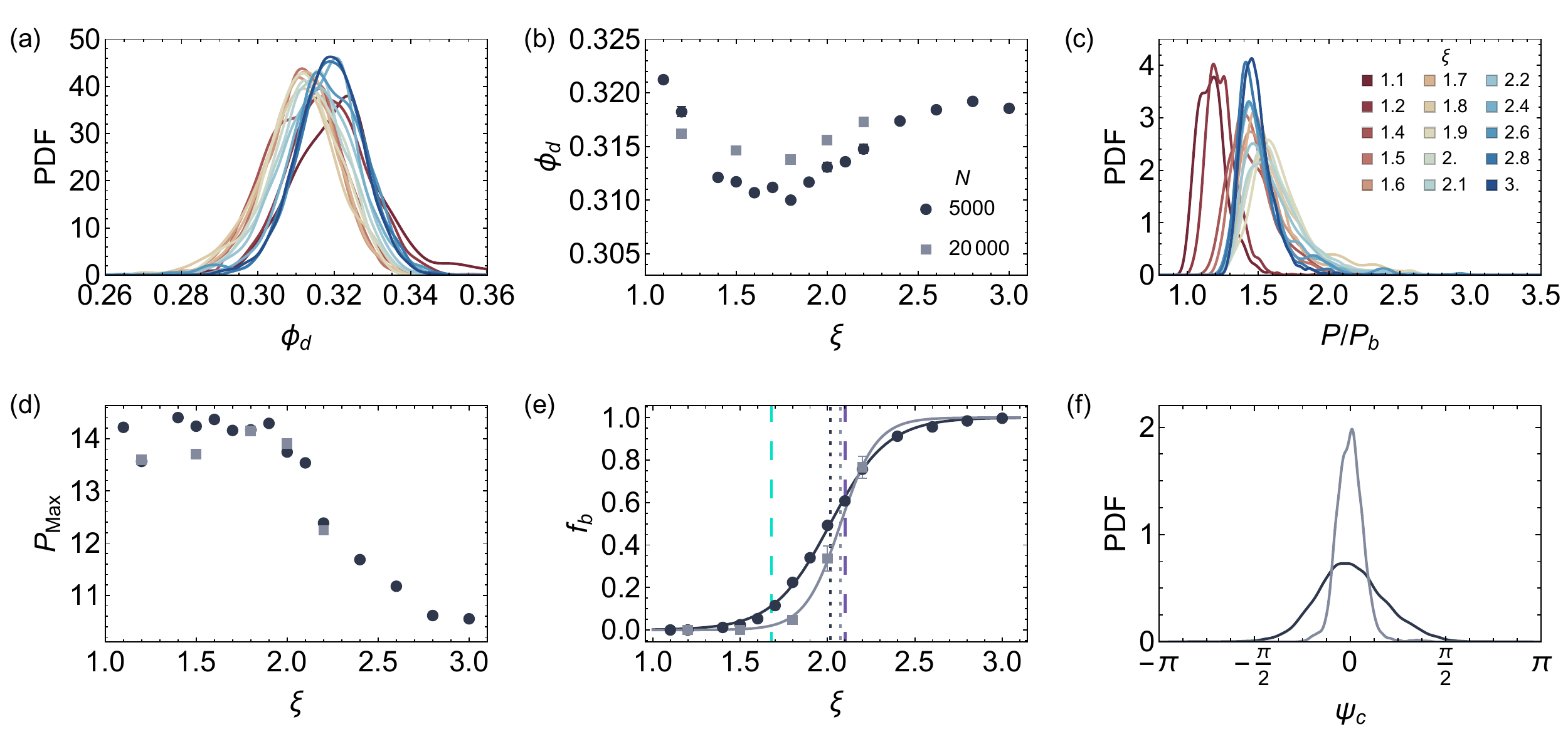}
      \caption{
      (a) Probability density distribution of the area fraction $\phi_d$ of the torus covered by the dense phase for different aspect ratios $\xi$. The color scale is shown in panel c.
      (b) Variation in the mean area fraction with aspect ratio for different numbers of particles $N$.
      (c) Probability density distribution of the perimeter $P$ of the dense phase, scaled by the perimeter of an ideal band $P_b$.
      (d) Peak value of the perimeter distribution for different $\xi$ and $N$.
      (e) Fraction of snapshots $f_b$ that form a band as $\xi$ is varied. The solid lines are fits to $f_b = \frac{1}{2}\tanh{\frac{\xi-\xi_c}{h}}+\frac{1}{2}$ for each $N$, which give $(\xi_c, h) = (2.01, 0.3)$ for $N=5000$ and $(2.07,0.2)$ for $N=20,000$.
      The $\xi_c$ values are indicated by the dotted lines while the turquoise and purple dashes respectively show predictions for $\xi_c$ for disks of constant geodesic curvature ($\xi_c=1.68$) and radius ($\xi_c=2.10$). 
      (f) Probability density distribution of the $\psi$ coordinate center of the dense phase for $\xi=1.5$ for different $N$.
    The outer equator of the torus has $\psi=0$.
    Distributions in panels a, c, and f are histograms constructed with a Gaussian smoothing kernel.
    The color scale in panel b also applies to panels d--f.}
    \label{fig:cluster-features}
\end{figure*}

To characterize features of the dense phase structure, we calculate its curved-space area and perimeter. 
Details of the calculations are given in the Appendix.
\Cref{fig:cluster-features}(a) shows the resulting distributions of the area fraction $\phi_d$ of the surface covered by dense phase as $\xi$ is varied.
We observe that the overall shapes of the distributions are approximately Gaussian for all $\xi$ values, and that the mean $\phi_d$ across all of the $N=5000$ data is $\phi_d^*=0.314$.
There is slight shift in the distributions as $\xi$ is varied, which we further investigate by plotting the mean area fraction of the dense phase for each $\xi$ in \cref{fig:cluster-features}(b).
This indicates that $\phi_d$ initially decreases as $\xi$ increases, before increasing again around $\xi=2$ to reach a plateau at large $\xi$.
This total change in the area fraction is small compared to the distribution width, and represents a change of $\sim 3\%$ of the total number of particles in the dense phase.
The effect also becomes even smaller at larger $N$, suggesting the slight variation could be due to the effect of curvature being greater at low $N$ near the inner equator, causing a slight reduction in cluster formation at intermediate $\xi$. 
Overall this result shows that the dense phase cluster area is largely independent of the torus aspect ratio at state points well within the phase separated regime.

By contrast, the cluster perimeters, calculated as the total geodesic length of the cluster boundaries, show a much stronger $\xi$ dependence.
The distributions, shown in \cref{fig:cluster-features}(c) with the perimeter $P$ scaled by the perimeter of an idealized banded configuration ($P_b = 2 C_r$, where $C_r= 2\pi /\sqrt{\xi}$ is the minor circumference of the torus), have a clear peak and a long tail.
At larger $\xi$, where the dense phase forms a band, the peaks of the distributions appear to converge to a value of $1.4 P_b$, indicating that in this regime the typical measured perimeter length relative to the minor circumference is constant.
Examining the position of the distribution peaks $P_{\text{Max}}$ in \cref{fig:cluster-features}(d), we find that the typical perimeter remains constant in the low-$\xi$ disk regime.
In the banded regime, $P_{\text{Max}}$ decreases with $\xi$ such that the ratio $P_{\text{Max}}/P_b$ plateaus.

We characterize the transition from a disk-shaped dense phase to a band by determining the fraction of banded snapshots $f_b$ at a given $\xi$, details of which are given in the Appendix. 
\Cref{fig:cluster-features}(e) shows how $f_b$ increases from zero at small $\xi$, and saturates at one for $\xi\gtrsim2.8$.
By fitting a curve of the form
\begin{equation}
f_b = \frac{1}{2}\tanh(\frac{\xi-\xi_c}{h}) + \frac{1}{2}
\end{equation}
we estimate that the critical aspect ratio for the transition is $\xi_c=2.02$ and the width of the transition is $h=0.3$ for $N=5000$.
At larger $N$, our select data points show that $\xi_c$ shifts to a larger value ($\xi_c=2.07$, $h=0.2$ for $N=20,000$), indicating that our estimate for $\xi_c$ should be taken as a lower bound on the true transition value.

Returning to the cluster density profile for low $\xi$ (\cref{fig:schematics-density}(c)), we observe that the resulting averaged cluster shape suggests that the cluster center is typically close to the outer equator of the torus, in the region of positive Gaussian curvature. 
This suggests that the cluster does not uniformly explore the entire surface, as it would in Euclidean space.
To investigate this further, we plot the distribution of the cluster $\psi$ coordinate center $\psi_c$ in \cref{fig:cluster-features}(f).
This confirms that the cluster centers are peaked around the outer equator, with very few clusters centered in the inner region of the torus ($\abs{\psi}>\pi/2$).
At larger $N$, the distribution becomes narrower, highlighting that the localization of the dense phase to the outer equator is robust to larger $N$ and is not merely a consequence of the potential reduced clustering in the higher curvature regime at the inner equator for $N=5000$.

\subsection{Thermodynamic and kinetic driving mechanisms}

What causes the cluster to preferentially sit at the outer equator, and its eventual transition to a band?
To answer this, we must examine the physical mechanisms governing the dense phase interface. 
In Euclidean space, the two dominant theoretical frameworks for MIPS yield identical predictions for the macroscopic cluster shape.
Thermodynamic models, based on capillary tension and interfacial dynamics, suggest that the cluster minimizes its boundary length for a given area, analogous to an equilibrium fluid droplet~\cite{patch2018curvature,langford2024theory}.
Alternatively, kinetic theories posit that the cluster is shaped by a local balance of particle deposition and escape rates, resulting in an interface that on average is equidistant from a central core~\cite{redner2013structure,soto2024kinetic}.
In flat space, both mechanisms produce a circular disk or a straight-edged band.

Curved space breaks this geometric degeneracy.
The thermodynamic framework of an emergent surface tension maps onto the mathematical isoperimetric problem; this is in general very challenging to solve, but the allowed solutions for a broad class of surfaces of revolution, including the torus, are known.
On the torus, isoperimetric domains are bounded by curves of constant \emph{geodesic curvature} $\kappa_g$~\cite{ritore2001constant,canete2007stable}, which quantifies the curvature of a curve $\gamma$ relative to the underlying surface---geodesics have $\kappa_g=0$, while nonzero $\kappa_g$ gives information about how $\gamma$ curves within the surface~\cite{kamien2002geometry}.
These domains are bands bounded by meridians (lines of constant $\theta$, which are geodesics and therefore have $\kappa_g=0$), and disks with constant $\kappa_g$ that are symmetric with respect to the equators~\cite{canete2007stable}.
Due to the Gaussian curvature profile of the torus, a constant $\kappa_g$ disk centered at the outer equator will always have a shorter perimeter than one enclosing the same area at the inner equator.
This directly aligns with our observation in \cref{fig:cluster-features}(f) that the low-$\xi$ dense phase preferentially sits at the outer equator.

Conversely, a naive application of the kinetic framework predicts that a stable cluster maintains an equal geodesic distance from its center.
In curved space, this corresponds to a curve of constant \emph{geodesic radius}, $r_g$ measured relative to a central point or line.
In regions of varying curvature, curves of constant $\kappa_g$ and constant $r_g$ take on distinct morphological profiles.
By analyzing the dense phase on the torus, we can therefore directly probe which geometric bound the MIPS cluster actually obeys.

To first determine the expected disk-to-band transition point, $\xi_c$, under the isoperimetric assumption, we investigate how the perimeter $P$ of a constant $\kappa_g$ disk centered at the outer equator varies with $\xi$.
Details of how the constant $\kappa_g$ disks are constructed are given in the Appendix.
The results, plotted in \cref{fig:const-kappag-expectations}(a), show that for small $\phi_d$ the ratio $P/P_b$ is always less than one, indicating that the disk solution is always optimal.
At larger $\phi_d$, there is a critical aspect ratio above which the enclosed region can reduce its total perimeter by transitioning from a disk into a band.

Using $\phi_d^*$ from our simulation data leads to a predicted value of $\xi_c=1.68$ above which the banded solution becomes optimal.
This value is notably smaller than $\xi_c=2.0-2.1$ that we observed in our simulation data. 
One reason for this apparent disagreement could be that even above the predicted $\xi_c$ the constant $\kappa_g$ disks are still local solutions to the isoperimetric problem, even though the banded solution is the global optimum.
Close to the transition, the boundary of the constant $\kappa_g$ disk is mostly in the regions with positive or zero Gaussian curvature, as shown in the inset in \cref{fig:const-kappag-expectations}(a).
As such, there is a finite energy barrier for the system to transition between the disk and band-shaped structures. 
As $\xi$ increases, the boundaries of the disk become closer to the inner equator, meaning that it is easier for fluctuations to allow the system to find the globally optimal banded state.

\begin{figure}
    \centering
    \includegraphics[width=\linewidth]{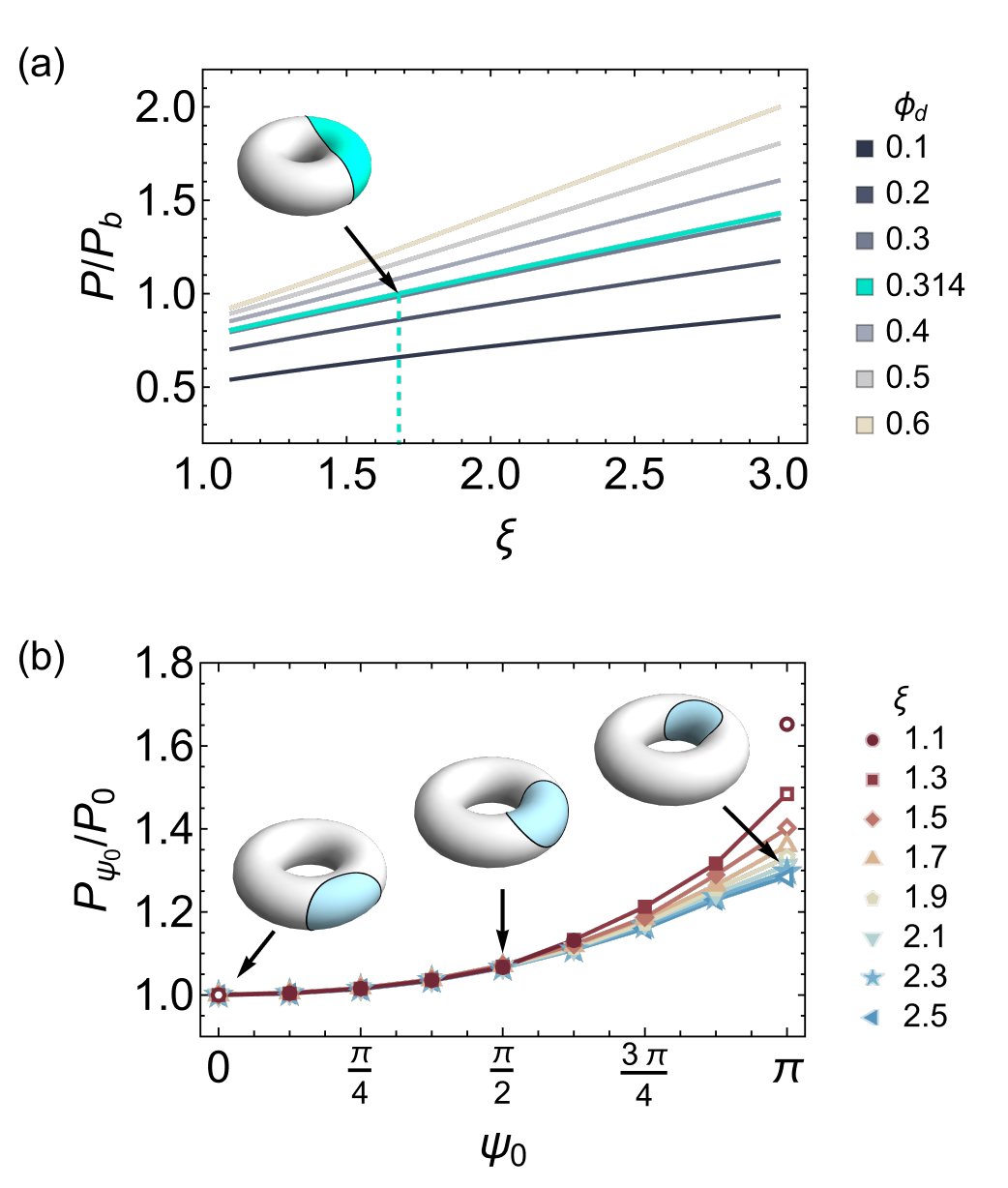}
    \caption{
    (a) Variation in the perimeter $P$ of a constant geodesic curvature $\kappa_g$ disk centered at the outer equator, scaled by the perimeter $P_b$ of an ideal banded solution as the aspect ratio $\xi$ is varied. 
    Curves correspond to different values of $\phi_d$, the area fraction covered by the disk. 
    The solid turquoise line indicates $\phi_d^*=0.314$, the average dense phase area fraction in the simulations, and the corresponding vertical dashed line shows the predicted transition value $\xi_c = 1.68$, above which a banded solution has a lower perimeter than a disk. 
    The torus inset shows the shape of the disk solution at the predicted $\xi_c$.
    (b) Variation in the minimized perimeter $P_{\psi_0}$ of a disk with fixed area fraction $\phi_d=0.1$ as the center position $\psi_0$ of the disk is varied for different values of $\xi$.
    Perimeter values are scaled by the perimeter $P_0$ of a disk centered at $\psi_0=0$. 
    Solid markers show solutions obtained via numerical optimization while open markers indicate values for which the solution is a curve of constant $\kappa_g$. 
    Note that for $\xi=1.1$, the numerical approach fails to find a valid solution near the inner equator. At the inner equator, the disk solution self-intersects slightly, indicating that the solution would instead be a cylinder spanning the inner equator.
    Inset tori show solutions for $\xi=2.1$ for $\psi_0=0,\frac{\pi}{2},\pi$.
    }
    \label{fig:const-kappag-expectations}
\end{figure}

Although the optimal solution for low $\xi$ is a disk centered at the outer equator, our simulation data indicates that the cluster center does deviate from this line (see \cref{fig:cluster-features}(f)). 
We therefore explore how moving the disk away from the equator changes the resulting boundary perimeter.
Given that the isoperimetric solutions are centered at the outer equator, these deviations give an indication of the energy penalty of the center straying from the outer equator.
Any disk centered away from the equator is, at best, a local solution to the isoperimetric problem.

To find these local solutions as the disk center location is varied, we use the shape optimization software Morpho~\cite{morpho2025} (see the Appendix for details).
We consider a disk with $\phi_d=0.1$ both for technical reasons in the numerical implementation and so that the disk is a valid solution close to the inner equator without self-intersecting.
\Cref{fig:const-kappag-expectations}(b) shows the disk perimeters $P_{\psi_0}$ as the center location $\psi_0$ is varied, plotted relative to the perimeter $P_0$ of the disk at the outer equator ($\psi_0=0$).
Note that the $P_0$ values at the outer equator are almost identical across $\xi$ values, with less than $2\%$ variation between the minimum and maximum value.
We observe that, as expected, the perimeter of the disk increases as the cluster center approaches the inner equator, with the $\xi$-dependent maximum perimeter being $1.3-1.65P_0$. 
Most of the increase in perimeter occurs in the region $\psi_0>\pi/2$, where the Gaussian curvature of the surface is negative. 
Up to $\psi_0=\pi/2$, the perimeter increase is only around $10\%$, meaning that deviations of the cluster center within the region of positive Gaussian curvature ($\psi<\pi/2$) give only a small penalty on the minimum possible length of the boundary compared to the optimal solution.

\begin{figure}
    \centering
    \includegraphics[width=\linewidth]{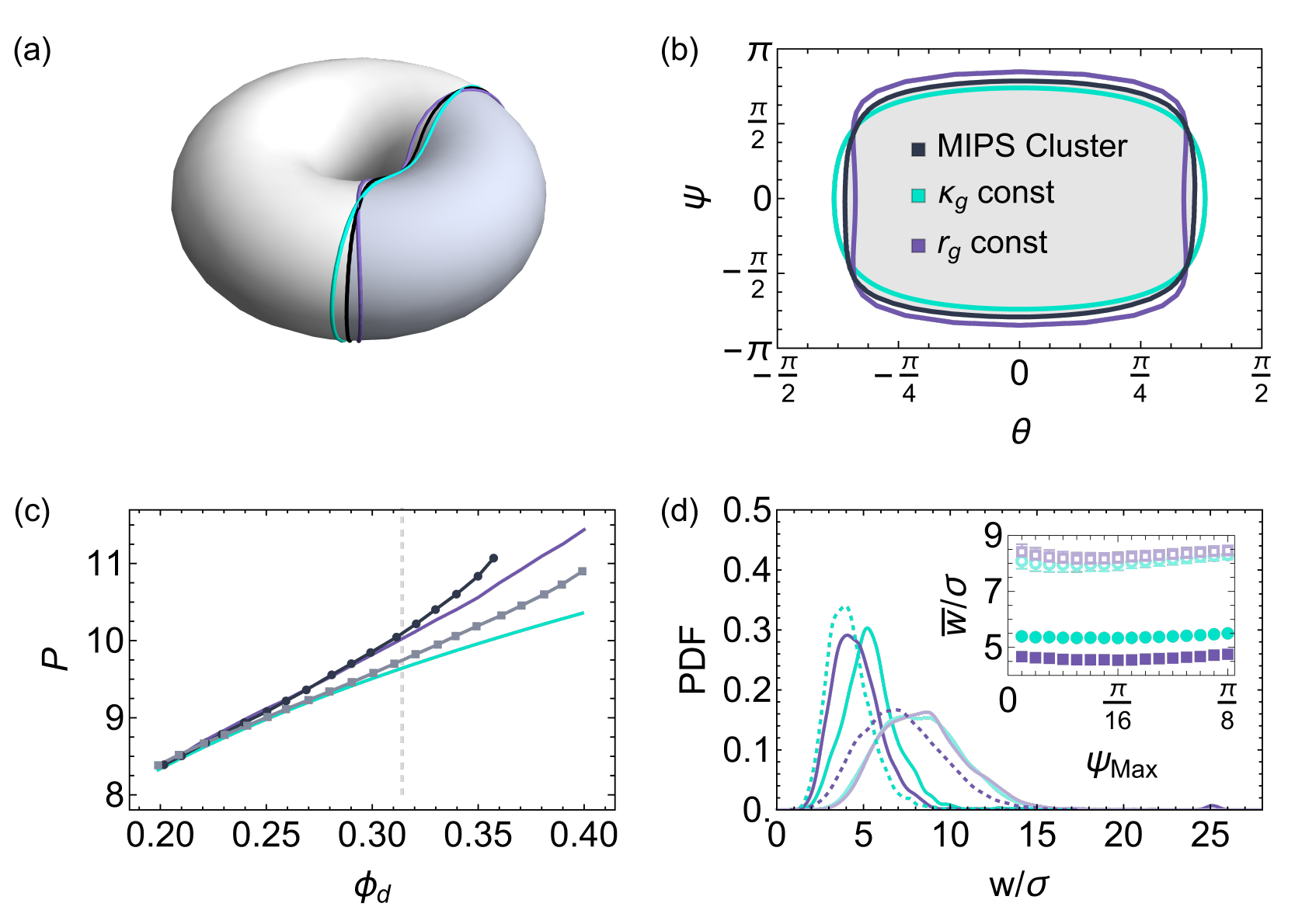}
    \caption{(a--b) Comparison between a contour of constant density, taken from the density plot in \cref{fig:schematics-density}(c) and area matched curves of constant geodesic curvature $\kappa_g$ and geodesic radius $r_g$ centered at the outer equator, shown (a) on the torus surface, and (b) in torus coordinate space.
    The area of the disks is taken as the average area fraction of the dense phase cluster $\phi_d^*=0.314$. 
    The color scale for the constant $\kappa_g$ and $r_g$ curves indicated in panel (b) applies to all figure panels.
    (c) Variation in the perimeter $P$ of these different curves as the area fraction $\phi_d$ is varied. Results are shown for contours taken from the $N=5000$ samples (black) and $N=20,000$ samples (gray). 
    The dashed vertical line indicates $\phi_d^*$.
    (d) Histogram of the perimeter deviation $w$ between individual disk-shaped clusters with centers within $\psi_{\text{Max}}$ of the outer equator and area-matched curves of constant $\kappa_g$ and $r_g$ centered at the outer equator. 
    Results are shown scaled by the particle diameter $\sigma$. 
    Solid lines are for $\xi=1.5$, $N=5000$, dashed lines are for $\xi=3.0$, $N=5000$ and lighter shades are for $\xi=1.5$, $N=20,000$. 
    (inset) Mean deviation $\bar w/\sigma$ for the $\xi=1.5$ distributions as the maximum allowed deviation of the cluster center from the outer equator is varied. 
    Solid markers are for $N=5000$, open markers are for $N=20,000$. 
    Error bars, which are typically smaller than the marker size, indicate the standard error across independent trajectories. 
        }
    \label{fig:cluster-deviations}
\end{figure}

To determine how the observed dense phase shape compares with theoretical expectations for disk-shaped clusters, we return to the averaged density profiles for $\xi=1.5$, shown in \cref{fig:schematics-density}(c) for $N=20,000$.
From these profiles, we extract contours of constant $\phi_d$, and compare their shape to area-matched regions bounded by curves with constant $\kappa_g$ and $r_g$ centered at the outer equator.
\Cref{fig:cluster-deviations}(a) shows such a comparison for a contour from the $N=20,000$ data enclosing an area fraction $\phi_d^*$. 
Qualitatively, the cluster contour is slightly closer in shape to the constant $\kappa_g$ solution than the constant $r_g$ one, although its shape sits between the two.

To quantify the differences between these shapes, in \cref{fig:cluster-deviations}(b) we plot the perimeters of each for a range of $\phi_d$ values.
At smaller $\phi_d$, the constant $\kappa_g$ and $r_g$ solutions are almost indistinguishable, which is to be expected for small regions close to the outer equator where the Gaussian curvature is approximately constant.
As $\phi_d$ increases and the constant $\kappa_g$ and $r_g$ solutions deviate, the perimeter of the constant $r_g$ solution becomes increasingly larger than the constant $\kappa_g$ solution.
Interestingly, we observe that the $N=5000$ contour perimeters track the constant $r_g$ perimeters very closely, while for $N=20,000$ the perimeters instead track the constant $\kappa_g$ perimeters for $\phi_d<\phi_d^*$. 
This therefore indicates that as the system approaches the continuum limit, the dense phase cluster shape appears consistent with the boundary-length-minimizing isoperimetric solution.

In addition to comparing the shapes of the averaged clusters, we can also compare the typical shapes of individual clusters to the two different types of geodesic disk.
We do this by measuring the averaged ``width'' $w$ of cluster interfaces, which quantifies the deviation between the cluster shape and area-matched disks with constant $\kappa_g$ and $r_g$.
We calculate $w$ using~\cite{patch2018curvature}
\begin{equation}
w^2 = \frac{1}{P_d}\int_0^{P_d} \dd s \abs{\delta h(s)}^2,
\label{eq:w2}
\end{equation}
where $s$ is a measure of arc length around the curve, $\delta h(s)$ is the distance between the reference curve and the interface, and $P_d$ is the total perimeter of the reference curve.
Further details of the calculation are given in the Appendix.

Here, we again focus on $\xi=1.5$, and consider only disk-shaped clusters that have their centers within an angle $\psi_{\text{Max}} = \pi/8$ of the outer equator. 
\Cref{fig:cluster-deviations}(d) shows the resulting histograms for $w$, measured in units of $\sigma$.
For $N=5000$, we observe that the constant $r_g$ histogram peaks at a smaller value than the constant $\kappa_g$ histogram, consistent with the observation that the averaged cluster data at this $N$ more closely tracks the constant $r_g$ data.
This is also reflected in the inset to \cref{fig:cluster-deviations}(d), which shows how the mean width $\bar w$ relative to the constant $r_g$ disks is consistently smaller as $\psi_{\text{Max}}$ is varied.
For $N=20,000$, we instead observe that the two histograms are very similar, but that the $\bar w$ values indicate that the typical cluster shape is slightly closer to that of a constant $\kappa_g$ disk.
Note that $\sigma$ for $N=5000$ is twice that for $N=20,000$, meaning that the actual $\bar w$ values are smaller for the higher $N$ in both cases.
These results are consistent with the averaged cluster shape analysis, again showing that the shape of the lower-$N$ data is closer to that of a constant $r_g$ disk, but at higher $N$ the shape becomes closer to the constant $\kappa_g$ disk.

In the banded regime, we find that for $\xi=3$ with $N=5000$ particles, the cluster shape is much more consistent with the isoperimeteric solution.
Using the averaged cluster density profile in \cref{fig:schematics-density}(d), we calculate the total perimeters of contours enclosing fixed $\phi_d$.
We observe that the contour perimeters remain roughly constant as $\phi_d$ is varied and that the contour perimeter $P_c$ enclosing area fraction $\phi_d^*$ is $1.02P_b$, where $P_b$ is the perimeter of the constant $\kappa_g$ banded solution.
By contrast, an area-matched region bounded by bands with constant $r_g$ has a perimeter $1.2P_c$.
Calculating $w/\sigma$ for these clusters (\cref{fig:cluster-deviations}(d)) also shows that the typical value for constant $\kappa_g$ is smaller than for constant $r_g$, with mean values $4.2\sigma$ and $7.3\sigma$ respectively.

\subsection{Geometric trapping on complex surfaces}

\begin{figure}
    \centering   
    \includegraphics[width=\linewidth]{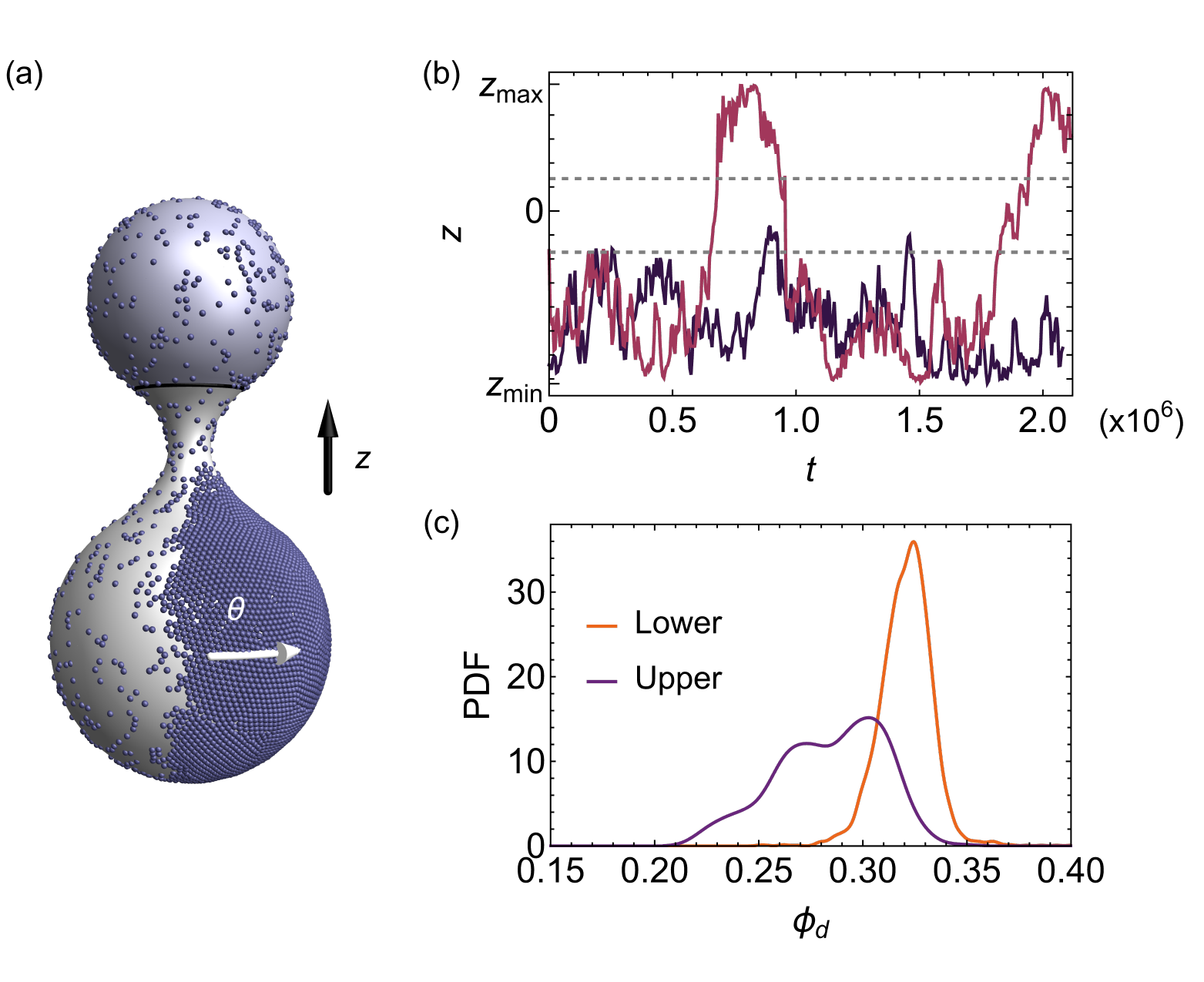}
    \caption{
        (a) Snapshot of the system on the hourglass surface with the coordinate directions marked. The shaded region represents the optimal isoperimetric solution on the surface for the expected dense phase area.
        (b) Time-dependence of the $z$-coordinate center of the dense phase cluster for two representative trajectories. In one trajectory, the dense cluster crosses between the lower and upper spheres; in the second, it remains centered on the lower sphere for the entire time sampled. Horizontal dashed lines indicate the boundaries of the neck region.
        (c) Probability distributions for the dense phase area fraction for clusters centered in different regions of the surface.
    }
    \label{fig:dumbbell}
\end{figure}

Having established that variable curvature on the torus dictates the steady-state shape of the dense phase while creating kinetic barriers to structural transitions, we next investigate whether these competing geometric principles can be leveraged to intentionally trap active clusters.
To test this, we consider an hourglass surface consisting of two spheres connected by a narrow neck, as shown in \cref{fig:dumbbell}(a).
This surface is constructed such that a dense phase covering the typical area fraction $\phi_d^*$ observed on the torus could fit entirely on the upper sphere, as indicated by the shaded region in \cref{fig:dumbbell}(a). 
Details of the surface and simulations are given in the Appendix.
On the hourglass, the isoperimetric solutions include regions bounded by meridians (lines of constant $z$), as well as all spherical caps with boundaries entirely within one of the two spheres.
The globally optimal location for a disk covering an area $\phi_d^*$ is therefore to form a spherical cap in the smaller, upper sphere, as this yields the shortest total boundary length.
Alternative solutions, such as a spherical cap on the larger lower sphere, possess perimeters that are at least twice that of the optimal solution.
Note that spherical caps on either sphere also have constant $r_g$.

To determine whether the dense phase localizes to this thermodynamically predicted location, we track the $z$-coordinate of the cluster center across trajectories (see \cref{fig:dumbbell}(b) for two representative trajectories).
Contrary to purely thermodynamic predictions, the dense phase does not preferentially localize to the upper sphere.
Instead, we observe that it spends the majority of its time in the lower sphere: in four of the ten recorded trajectories, the dense phase remains in the lower sphere for the entire $200,000\tau$ simulation window, and in the remaining trajectories the cluster transitions between the spheres.
In total, roughly $80\%$ of our snapshots have the cluster center below $z=0$, despite this lower region accounting for only $65\%$ of the total surface area.
The narrow, negative-curvature neck appears to act as a substantial bottleneck for the cluster crossing between the two spheres. 
When the cluster does cross between spheres, the transition through the neck region occurs over relatively short timescales compared to the time the cluster spends exploring either of the two spheres.
These trajectories indicate that the surface effectively creates a two-level system, where the cluster is almost always centered on one sphere or the other for a sustained duration, separated by an effective energy barrier to crossing.
We also note that while trajectory data indicates that transitions between spheres are relatively quick, statistical sampling reveals the cluster center occasionally lingers in the neck region, further highlighting how this narrow constriction acts as a kinetic barrier to the cluster transferring between the two spheres.

To understand why the dense phase dynamics deviate from the purely shape-based isoperimetric predictions, we examine the area fraction of the dense phase on the upper and lower spheres.
We include only clusters centered above or below the respective sphere equators to avoid the immediate influence of the neck region.
As shown in \cref{fig:dumbbell}(c), we find that the area distribution for the lower sphere clusters is consistent with observations on the torus, with a mean value $\phi_d=0.32$.
On the upper sphere, however, the distribution is much broader, and the mean shifts to $\phi_d=0.28$.
We also observe this distribution is sensitive to the choice of center cutoff, with the smaller peak disappearing and the distribution becoming narrower when the cutoff is set further up the upper sphere. 
This suggests that the dense phase is unable to easily maintain its preferred size on the upper sphere, likely because the nearby high-negative-curvature neck region prevents enough particles from remaining in the cluster. 
These observations demonstrate how highly localized curvature regimes, in conjunction with global shape-based considerations, create complex energetic landscapes that can trap and localize the dense phase.

\section{Discussion}

Our results demonstrate that spatial curvature exerts robust geometric control over the dense phase in motility-induced phase separation, even in parameter regimes where the scale of the Gaussian curvature relative to the particle size ($\abs{K}\sigma^2$) is sufficiently small that the overall phase boundaries remain largely unaffected~\cite{schonhofer2022curvature}.
While the total area of the dense phase remains independent of the underlying geometry, its morphology and spatial localization are dictated by the surface. 
By leveraging the distinct geometric properties of curved space, we explored the structural degeneracy present in flat-space MIPS and differing theoretical frameworks governing the dense phase.

Our numerical results and analysis reveal an interesting tension.
The steady-state preferred locations of the dense clusters on low-aspect-ratio tori align very well with thermodynamic, surface-tension-driven models~\cite{patch2018curvature,langford2024theory}. 
The shape of these dense-phase disks is more ambiguous: simulations of larger numbers of particles conform more closely with the thermodynamic picture, whereas those with fewer particles align better with the kinetic framework (see \cref{fig:cluster-deviations}(c-d)).
By contrast, in the banded regime the cluster shape is consistent with the thermodynamic picture even at the lower number of particles. 
Our results therefore point towards the cluster shape being ultimately consistent with expectations for thermodynamic, surface-tension-driven shapes in the limit of vanishing curvature (which our larger-$N$ simulations tend toward).

However, while the steady-state \emph{shape} may be controlled by effective surface tensions, our data suggests that the \emph{transition} between morphologies is kinetically limited. 
The shift from a localized disk to a continuous band occurs at an aspect ratio ($\xi_c \approx 2.0-2.1$) that is significantly larger than the purely isoperimetric prediction ($\xi_c = 1.68$). 
Notably, our simulations with increasing $N$ indicate a shift of the transition point \emph{even farther} from the isoperimetric prediction.
We hypothesize that this discrepancy arises from a substantial geometric energy barrier.
At the predicted $\xi_c$, transitioning from the optimal disk to a band would require massive interfacial fluctuations. 
Interestingly, calculating the aspect ratio at which a purely kinetic, constant $r_g$ disk of area fraction $\phi_d^*$ self-intersects yields $\xi_c = 2.10$, closely mirroring our observed transition point. 
This implies that while the cluster strives for an isoperimetric shape, the physical transition to a banded state is dictated by deposition-escape kinetics and the necessity for the fluctuating cluster boundary to self-intersect.
We note that while we clearly observe a structural transition on the torus, to our knowledge the equivalent critical aspect ratio for a disk-to-band transition in a flat, periodic system is currently unknown.
Characterizing this flat-space counterpart will be a valuable next step to fully isolate the influence of the curved manifold from the effects of the periodic boundary conditions.

This interplay between preferred thermodynamic shapes and kinetic barriers is further reinforced by our results on the hourglass manifold. 
Although the global isoperimetric optimum lies on the smaller upper sphere, the dense phase is predominantly trapped on the larger \emph{lower} sphere. 
The narrow, negative-curvature neck acts as a kinetic bottleneck, disrupting the local deposition balance and preventing the cluster from reaching its global geometric minimum.
This likely stems from the neck region falling outside the low-curvature regime identified in Ref.~\cite{schonhofer2022curvature}, meaning the local phase boundary for MIPS is expected to be shifted in this region.
This suggests further simulations at larger system sizes may be needed to determine whether these observations prevail in lower curvature regimes.

Together, these findings highlight how geometry can serve as a powerful tool for probing the physics of non-equilibrium systems, in this case by breaking degeneracies that exist in flat space. 
Looking forward, the coupling between curvature-dependent phase behavior, isoperimetric shape preferences, and kinetic bottlenecks opens the door to another flavor of programmable active matter.
By designing complex surfaces with specific curvature gradients and topological bottlenecks, it should be possible to engineer stable traps, guide optimal trajectories, and dictate the macro-scale morphology of active clusters. 
Investigating whether similar geometric control mechanisms are exploited in biological active systems, such as in the curvotaxis of cell collectives or the formation of curved biofilms, remains an exciting future direction.

\section*{Acknowledgements}
We thank Tim Atherton and Philipp Sch\"onh\"ofer for useful discussions.
HSA acknowledges funding from the Tarbutton Postdoctoral Fellowship.
This material is based upon work supported by the National Science Foundation under Grant No. DMR-2143815.

\appendix
\section{Analysis on the torus}
\label{sec:appendix}

\subsection{Identifying clusters and bands}
\label{sec:clusters-bands}
We identify the dense phase cluster by constructing a nearest neighbor graph for the particles in $\mathbb{R}^3$ using a Euclidean distance cutoff of $1.1\sigma$. 
The value of $\sigma$ means that the Euclidean separation is a very good approximation for the geodesic separation.
To determine whether the cluster is a disk or a band, we first map the edges in the largest cluster graph into toroidal coordinate space $(\theta,\psi)$ and construct the eight closest periodic images of the cluster edges.
If the largest connected component(s) of the graph of the original cluster with all the images included is the same size as the original cluster, the cluster is a disk; otherwise it must be a band.
Here our focus is primarily on banding around the $\psi$ direction (minor circumference), although bands that span the $\theta$ direction at the inner equator sometimes emerge at small $\xi$.
We therefore record separately the direction in which the bands form.

To compare the averaged cluster shape across snapshots, we center each of the clusters in the $\theta$ direction.
In curved space, the center of an object can be interpreted in different ways.
Here we treat the center as the center calculated in coordinate space.

\begin{table}[t]
    \centering
   \begin{tabular}{|c c c c c|}
        \hline
         $N$ & $\xi$ & $N_s$ & $N_f$ & $\tau_f/\tau$\\
         \hline
	5000 & 1.5 & 20 & 200/450* & 200\\
	 & 1.6, 1.7, 2.1 & 10 & 450 & 200 \\
	 & 1.2, 2.0, 2.2 & 5 & 450 & 200 \\
	 & 3.0 & 10 & 540 & 400 \\
	 &1.1, 1.4, 1.8, 1.9, 2.4, 2.6, 2.8 & 1 & 250 & 200 \\
	  \hline
	 20000 & 1.2 &  1 & 450 & 300\\
	  & 1.5, 2.0, 2.2 & 10 & 200 & 300 \\
	  & 1.8 & 1 & 276 & 200 \\
	 \hline
    \end{tabular}
    \caption{Summary of data used in analysis of different state points on the torus. Here $N_s$ is the number of independent samples, $N_f$ is the minimum number of frames recorded from a given sample after reaching steady state, and $\tau_f$ is the time interval between recorded frames. *10 samples have 200 frames, 10 samples have 450 frames.
    }
    \label{tab:raw-data}
\end{table}

\subsection{Dense phase area and perimeter}
\label{sec:area-perim}
The area of a cluster on the torus is given by 
\begin{equation}
A = \int_{\text{cluster}} r (R+r \cos{\psi}) \dd\theta\dd\psi.
\end{equation}
To calculate this in our system, we construct a concave hull mesh for the centered cluster in torus coordinate space.
For disk-shaped clusters, the integration is performed over the region defined by the mesh.
If the cluster is a band, the mesh is constructed to include image clusters to eliminate artifacts at the periodic boundaries.
The region of integration is then the intersection of the mesh with the primary cell.

The cluster perimeter is calculated by identifying edges on the mesh boundary and calculating the total geodesic lengths of all the edges.
For bands, only the edges within the primary cell are included. 
For small separations, as is the case here for neighboring particles on the edge of the mesh, we use the approximate geodesic separation $\Delta s$ to speed up calculations.
For two points with coordinates $(\theta_1,\psi_1)$ and $(\theta_2,\psi_2)$, this is given by
\begin{equation}
\Delta s \approx \sqrt{ r^2 \Delta\psi^2 + (R+r \cos\psi_{\text{av}})^2\Delta\theta^2},
\end{equation}
where $\Delta\theta = \theta_2-\theta_1$, $\Delta\psi = \psi_2-\psi_1$ and $\psi_{\text{av}} = (\psi_1+\psi_2)/2$. 

We note that in this approach, the exact numerical values of area and perimeter depend on choices of the distance cutoff threshold for connecting points in the concave hull mesh, with higher thresholds leading to a cluster with smoother edges and a slightly larger enclosed area.
In torus coordinate space, particles at the inner equator of the torus have a larger separation than those at the outer equator. 
We therefore choose the cutoff threshold for the mesh to be as small as possible while ensuring all of the particles in the cluster are included within the mesh.
At smaller $\xi$, where this effect is most pronounced, this may lead to a slight over-estimation of the cluster area and underestimation of the perimeter as the resulting mesh is slightly smoothed to incorporate points near the inner equator.
However, we believe that the overall trends reported here are robust to reasonable small changes in this threshold.

\subsection{Using Morpho for numerical optimization}
\label{sec:morpho}
We use the open-source shape optimization software  \emph{Morpho}~\cite{morpho2025} to find the isoperimetric solution for a disk centered at an initial location $(\theta_0,\psi_0)$ on the torus. 
In the problem setup, the disk is a circular 2D mesh with a prescribed area $A_d = \phi_d A_T$, where $\phi_d=0.1$ is the area fraction and $A_T$ is the final surface area of the torus.
The optimization minimizes the perimeter of the disk subject to the constraint $A_d$ is fixed.
The disk is constrained to the torus surface through a level-set constraint that imposes an energy penalty for deviation from the surface.

In the initial setup, the disk is placed at a chosen starting location with its center in contact with the torus surface and the disk in the local tangent plane to the torus.
To allow the disk to wrap onto the torus properly, we initially use a torus with minor radius $r=2r_d$, where $r_d$ is the initial radius of the disk.
During optimization, we iteratively allow the disk to find its preferred shape before reducing the torus size at fixed aspect ratio.
Here we perform this optimization procedure five times. 

Using a relatively small $\phi_d$ means that the disk remains centered around its initial location, and we are able to more carefully track how the local curvature variation changes the resulting perimeter.
Note that for small $\xi$, the optimal solution near the inner equator is topologically a cylinder, not a disk. 
We therefore exclude points where the disk overlapped from the plot in \cref{fig:const-kappag-expectations}.

\subsection{Curves of constant geodesic curvature}
\label{sec:kg-curves}
Consider a curve on the torus $\gamma(s) = (\theta(s),\psi(s))$, parameterized by arc length $s$.
Using the parameterization of the torus in \cref{eq:r-torus}, we can define a set of orthonormal basis vectors $\vu{e}_{\theta}$, $\vu{e}_{\psi}$ that live in the tangent plane to the surface at each point and are parallel to the two coordinate directions. The surface normal $\vu{e}_n = \vu{e}_{\theta}\times\vu{e}_{\psi}$ gives us a complete basis.
The tangent vector $\vu{T}=\dv{\vb{r}(s)}{s}$ to $\gamma(s)$ can then be written as
\begin{equation}
\vu{T} = \cos{\beta(s)} \vu{e}_{\psi} + \sin{\beta(s)}\vu{e}_{\theta},
\end{equation}
where we have introduced $\beta(s)$ as the angle between the curve and the $\vu{e}_{\psi}$ direction.
This obeys
\begin{align}
\cos{\beta(s)} &= r \dv{\psi(s)}{s} \\
\sin{\beta(s)} &= \left(R+r\cos{\psi(s)}\right)\dv{\theta(s)}{s}.
\end{align}

The geodesic curvature of the curve is given by, see e.g. Ref.~\cite{kamien2002geometry}
\begin{equation}
\kappa_g(s) = \vb{T}'(s)\vdot\left[\vu{e}_n(s)\times \vu{T}(s)\right],
\end{equation}
where $\vb{T}'(s) = \dv{\vb{T}}{s} = \kappa(s)\vb{N}(s)$, $\vb{N}(s)$ is the unit normal to $\gamma(s)$ and $\kappa(s)$ is the curvature of $\gamma(s)$.
For our system, this gives
\begin{align}
\kappa_g(s) &= \beta'(s) - \sin{\psi(s)}\theta'(s) \\
&= \beta'(s) - \frac{\sin{\psi(s)}}{R+r\cos{\psi(s)}}\sin{\beta(s)}.
\label{eq:kappa_g}
\end{align}
Curves with a constant $\kappa_g$ can therefore be constructed by numerically solving \cref{eq:kappa_g} with $\kappa_g$ prescribed.

Here, our goal is to determine the value of $\kappa_g$ that encloses a fixed area.
Starting from an initial guess for $\kappa_g$, we use Stokes' theorem to calculate the area of the resulting disk. 
We then use interval bisection to refine this estimate of $\kappa_g$ until the resulting disk area is within a prescribed tolerance of the target area. 
The perimeter of the disk is determined as the point where the numerical solution self-intersects.

\subsection{Curves of constant geodesic radius $r_g$}
To construct curves of constant geodesic radius, we first find a geodesic $\gamma(t)$ that starts from the chosen center $(\theta_0,\psi_0)$ at an initial angle $\beta_0$, measured relative to the $\vu{e}_{\psi}$ direction.
The resulting geodesic equations obeyed by $\gamma(t) = (\theta(t),\psi(t))$ are
\begin{align}
\ddot\psi(t) + \frac{(R+r\cos{\psi(t)})\sin{\psi(t)}}{r}\dot\theta(t)^2 &= 0 \label{eq:geopsi} \\
\ddot\theta(t) - \frac{2r\sin{\psi(t)}}{R+r\cos{\psi(t)}} \dot\theta(t)\dot\psi(t)&= 0. \label{eq:geophi} 
\end{align}
We set the initial conditions 
 \begin{align}
 \psi(0) = \psi_0, &\quad \theta(0) = \theta_0, \quad \\
 \dot\psi(0) = \frac{\cos{\beta_0}}{r}, &\quad \dot\theta(0)=\frac{\sin{\beta_0}}{R+r \cos{\psi_0}},
 \label{eq:bcs}
 \end{align}
 such that $\gamma(t)$ has unit speed parameterization. 
 
We numerically solve these equations and record the end position after traveling the prescribed geodesic radius $r_g$ along $\gamma(t)$. 
To construct the full constant $r_g$ circle, we repeat this procedure for a range of $\beta_0$ values spanning the entire circle (typically 60 or 120 values).
The perimeter of the circle is calculated as the sum of the geodesic separations between neighboring pairs of sampled end points, and the area is numerically integrated.
To find a constant $r_g$ disk with a prescribed area, we start with an initial guess for $r_g$ and calculate its area.
We then use interval bisection to refine the estimate until it is within a prescribed tolerance of the target area.

\subsection{Cluster interface width}
\label{sec:cluster-interface}
We calculate the mean interface width $w$ of individual clusters using \cref{eq:w2}.
For disk-shaped clusters ($\xi=1.5$), we focus on clusters that are centered close to the outer equator, and consider all clusters for which the coordinate center $\psi_c$ obeys $\psi_c\leq \psi_{\text{Max}}$, with $\psi_{\text{Max}}=\pi/8$.
This corresponds to 52\% of snapshots for $N=5000$ and 94\% for $N=20,000$. 

To find $w$ for a given snapshot, we calculate the area of the dense phase cluster, then construct the corresponding constant $\kappa_g$ and $r_g$ solutions with their $\theta$-coordinate centers aligned.
We then sample points along the perimeter of each of the constant $\kappa_g$ and $r_g$ curves, and calculate the geodesic distance between each curve and the edge of the cluster.
This is calculated as the geodesic distance traveled along a geodesic ray emitted perpendicular to the reference curve at a given point.
For constant $\kappa_g$, points are sampled evenly along the arc length of the curve.
For constant $r_g$ disks, points are sampled at equally spaced angles measured from the center of the disk, while for the constant $r_g$ bands points are sampled evenly around the $\psi$-direction.
In each case, at least 60 points around the reference curves are used to calculate $w$.

\section{Hourglass surface}
\label{app:dumbbell}
The hourglass surface consists of two spheres with different radii and offset centers that are connected using a cubic spline that ensures the surface is smooth at the point of connection.
In cylindrical coordinates $(r, \theta, z)$, the radius of the surface at a given $z$ is given by 
\begin{equation}
r  = 	\begin{cases}
	\sqrt{R_U^2-(z-c_U)^2} &\text{if } z_U \leq z \leq z_{\text{max}} \\
	a z^3 + b z^2 + c z + d &\text{if } z_L< z < z_U\\ 
	\sqrt{R_L^2-(z+c_L)^2} &\text{if } z_{\text{min}} \leq z \leq z_L, \\
	\end{cases}
\end{equation}
where $R_U=3$ and $R_L=2.2$ are the radii of the upper and lower spheres, $c_U=h R_U$ and $c_L=h R_L$ with $h=1.4$ are the center locations of the two spheres along the $z$-axis, and the polynomial coefficients $a\approx 0.002$, $b\approx 0.45$, $c\approx 0.06$, and $d \approx 0.46$ are numerically determined to ensure the surface is smooth.

This construction means that the upper portion of the surface ($z>0$) accounts for $35\%$ of the surface area, meaning that a dense phase cluster at the typical size observed in the torus would fit entirely in this region.
Here, we use $N=5000$ particles with $\phi=0.4$. 
The curvature scales on the lower and upper spheres are $K\sigma^2 = 0.002$ and $K\sigma^2 = 0.0038$ respectively.
The lower sphere is therefore in the ``low curvature'' regime identified in Ref.~\cite{schonhofer2022curvature}, meaning curvature has minimal effect on phase behavior, while the upper sphere is at the lower end of the ``intermediate curvature'' regime, meaning the positive curvature may slightly promote clustering.
The neck region has negative Gaussian curvature with a minimum value of $K\sigma^2 = -0.036$, suggesting this region would reduce clustering~\cite{schonhofer2022curvature}.

The $z$-coordinate center of the clusters is calculated by projecting the $\mathbb{R}^3$ coordinate center of the cluster onto the closest point on the surface.
For a spherical cap, this approach exactly aligns with the coordinate center of the cluster, which is equidistant from all points on the cluster boundary.
For our clusters, this approach provides a good approximation for the center, especially when the center is on one of the two spheres and away from the neck region.
In the neck region, alternative methods for determining the cluster center could yield slightly different coordinate values.
However, we find this approach gives a fast method for calculating the center that provides reasonable values across the entire surface.

\end{document}